# Microparticles self-assembly induced by travelling surface acoustic waves

Ghulam Destgeer,[1] Ali Hashmi,[2] Jinsoo Park,[1] Husnain Ahmed,[1] Muhammad Afzal,[1] and Hyung Jin Sung[1,*]

[1]Department of Mechanical Engineering, KAIST, Daejeon 34141, Korea.
[2]Institut de Biologie du Développement de Marseille (IBDM), France.

We present an acoustofluidic method based on travelling surface acoustic waves (TSAWs) for the induction of the self-assembly of microparticles inside a microfluidic channel. The particles are trapped above an interdigitated transducer, placed directly beneath the microchannel, by the TSAW-based direct acoustic radiation force (ARF). This approach was applied to 10 µm polystyrene particles, which were pushed towards the ceiling of the microchannel by 72 MHz TSAWs to form single- and multiple-layer colloidal structures. The repair of cracks and defects within the crystal lattice occurs as part of the self-assembly process. The sample flow through the first inlet can be switched with a buffer flow through a second inlet to control the number of particles in the crystalline structure. The constant flow-induced Stokes drag force on the parti-cles is balanced by the opposing TSAW-based ARF. This force balance is essential for the acoustics-based self-assembly of microparticles inside the microchannel. Moreover, we studied the effects of varying the input voltage and fluid flow rate on the position and shape of the colloidal structure. The active self-assembly of microparticles into crystals with multiple layers can be used in the bottom-up fabrication of colloidal structures with dimensions greater than 500 µm x 500 µm, which is expected to have important applications in various fields.

## Introduction

The self-assembly of individual microparticles into ordered crystals[1] is important for the bottom-up fabrication of supra-colloidal microstructures,[2] which are essential to the fields of photonics,[3] sensors,[4] robotics,[5] and microfluidics,[6] among others.[7] External magnetic[8,9] and electrodynamic[10,11] force fields have been **used** to facilitate the self-assembly process and found to reduce the timescale of self-assembly and to provide better control over particle motion. However, these techniques are dependent upon the electrical and magnetic properties of the particles and the suspension media, which limits their applicability.[12–14] Acoustic forces have been used to manipulate (align,[15–17] pattern,[18–20] separate,[21–23] and concentrate[24–26]) microparticles based on their size differences and mechanical properties. Acoustic waves are suitable for the handling of virtually any micro-object and are not dependent on the electro-magnetic properties of the material and the suspension media. Acoustic waves have recently been used in the manipulation of colloidal particles and their self-assembly.[27–30]

Owens et al.[28] have demonstrated the assembly of particles into ordered colloidal crystallites by using standing bulk acoustic waves in a batch process. They employed two lead zirconate titanate (PZT) transducers to generate standing waves within a reflective microchamber and thus to concentrate particles at the acoustic pressure nodes in the two-dimensional plane. Akella and Juárez[30] employed a continuous flow acoustofluidic device with a single PZT element to generate a region of low acoustic pressure within a microfluidic channel in order to concentrate and assemble particles flowing through the acoustic field region. Yang et al.[31] used acoustic and magnetic fields in combination to control the particle concentration and inter-particle interactions respectively and thus to fabricate a range of colloidal structures. Most research to date on microparticle self-assembly with bulk acoustic waves has been based on the generation of acoustic pressure nodes in a standing wave. The size of the colloidal crystals formed with this approach is limited by the saturation limit of the pressure nodes. Wollmann et al.[32] utilized travelling surface acoustic waves (TSAWs) with frequencies two orders of magnitude higher than those of techniques using bulk acoustic waves to direct the self-assembly of colloidal crystals in an evaporating sessile droplet. Similarly, in a recent study, the separation of microparticles inside an evaporating sessile droplet has been demonstrated, although colloidal crystal formation was not explicitly studied.[21] The separation of particles inside sessile droplets was achieved by making use of the difference between the acoustic radiation forces (ARFs) acting on particles of different sizes,[21] in contrast to the induction of crystal growth within a sessile droplet dominated by an acoustic streaming flow (ASF) and standing waves, formed due to the internal reflection.[25,32]

In this article, we demonstrate a TSAW-based technique for the self-assembly of microparticles flowing inside a microfluidic channel, in which the particles are trapped above an interdigitated transducer (IDT) by a direct ARF (see Figure 1). A polydimethylsiloxane (PDMS) microfluidic channel is placed directly on top of an IDT in order to couple acoustic waves directly into the fluid. Single or multiple layers of colloidal crystals self-assemble because the ARF on the particles is balanced by the flow-induced drag force. The size and number of layers of the crystalline structure can be controlled by mediating the flow of particles into the trapping zone, which is achieved by switching the sample flow through the first inlet with a buffer flow through a second inlet. The crystalline structure was studied for periods of several minutes up to an hour in order to track crack propagation and self-repair in the colloidal crystal. The proposed TSAW-based technique circumvents the saturation limits of standing-wave pressure nodes[28,30] as well as the drawbacks of sample recovery in sessile-droplet-based systems.[21,32] The effects of varying the input voltage and the flow rate on the development of the crystalline structure within the microchannel were also studied.

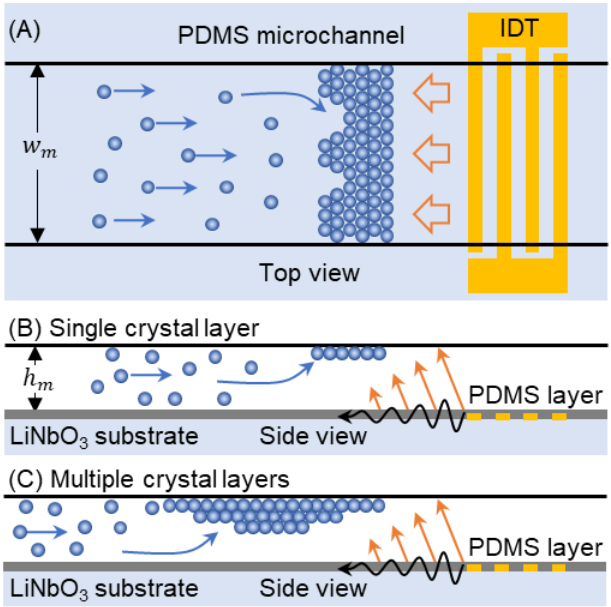

**Figure 1.** Schematic diagram of a TSAW-based trapping platform for microparticle self-assembly. (A) Top view: the TSAWs originate from an interdigitated transducer (IDT) and stop microparticles flowing inside a polydimethylsiloxane (PDMS) microchannel (width, $w_m$). (B) Side view: the acoustic waves push the particles up towards the ceiling of the microchannel (height, $h_m$). A single-layer crystal is formed through particle self-assembly. A thin PDMS sealing layer separates the IDT from the fluid inside the microchannel. (C) The continuous flow of particles to the trapping zone leads to the formation of a multiple-layer crystal.

## Methods

**Device fabrication:** A single-layered PDMS microchannel ($w_m \times h_m$: 500 μm × 110 μm) was fabricated by using the commonly used mold replica soft lithography process.[33–35] After cutting the desired PDMS microchannel, two inlet ports and one outlet port were punched through (Harris Uni-Core, I.D. 1 mm) the PDMS. A thin layer of PDMS solution (thickness, $t$ = 38 μm) was spin-coated onto a salinized Si wafer to fabricate a sealing membrane. After curing at 65°C for more than an hour, the PDMS microchannel was attached to the thin membrane by performing oxygen plasma bonding (Covance, Femto Science, Korea). A pair of uniformly spaced electrodes (IDT) was deposited parallel to the principal axis of a piezoelectric wafer (lithium niobate, LiNbO$_3$, 128 Y-X cut, MTI Korea) by using E-beam evaporation and lift-off processes (IAMD, SNU, Korea). The IDT consists of 20 electrode finger pairs with a 1 mm aperture, has an actuation frequency of 72 MHz and an output with a wavelength ($\lambda$) of 50 μm, and is covered with a thin layer of SiO$_2$ (2000 Å) to protect it from mechanical damage. The thickness ($t$) of the sealing PDMS membrane was chosen so that $t < 2\lambda$, which ensures that the ARF effect dominates the acousto-thermal effect inside the microchannel.[34,36] The microchannel was placed directly on top of the IDT to reversibly bind it to the piezoelectric substrate. Therefore, a single IDT can be used with various disposable PDMS microchannels.

**Experimental set-up:** A buffer solution was prepared by mixing deionized (DI) water and D$_2$O (Sigma Aldrich) in a volume ratio of 1:1 to obtain a solution density of 1.05 g/cm$^3$, and 1% surfactant (unless otherwise mentioned) was also added. Polystyrene particles (10 μm, NIST standard, density 1.05 g/cm$^3$, Thermo Scientific) were suspended in a density-matched buffer solution to prepare the sample. Two syringe pump units (neMESYS, Cetoni GmbH) were used to inject the buffer and sample solutions through the separate inlet ports. The IDT was actuated with an amplified (LZT-22+, Mini-Circuits) high frequency AC signal, produced by an RF signal generator (N5181A, Agilent Technologies). The device was mounted on a microscope (BX53, Olympus), while images were captured by using a CCD camera (DP72, Olympus) and analyzed with ImageJ software (http://imagej.nih.gov/ij/). For better visibility of and contrast between the microparticles, background noise was removed from the experimental images by using ImageJ (see Fig. 2, Fig. 3(A), Figs. 4(A)-(B)). In the experimental images, a single particle appears as a black disc with a bright spot at the center (see Fig. 3(B)); however, when the background noise is subtracted from the images, the particle appears as a white disc with a black spot at the center (see Fig. 3(A)).

**Working mechanism:** The microchannel is placed above the TSAW device so that the IDT covers the full width of the microchannel (see Fig. 1(A)). The particles pumped through the microchannel are stopped by the SAW when the Stokes drag force ($F_d$) acting on the particles is balanced by the ARF ($F_{ARF}$). The two forces, $F_{ARF}$ and $F_d$, determine the final position of the individual particles and of the particle clusters inside the microchannel as the crystal grows with the addition of incoming particles. The polystyrene particle diameter ($d_p$) and TSAW frequency ($f$) must be chosen to ensure that the Helmholtz number ($\kappa = \pi f d_p / c_f$) is greater than one, where $c_f$ is the speed of sound in the fluid, which means that the particles will experience a strong ARF. As the TSAW attenuates away from the IDT and couples with the fluid inside the microchannel, the compressional TSAW is radiated at an angle with a strong vertical ARF component.[17,24,33] Therefore, the particles are pushed up against the ceiling of the microchannel (see Fig. 1(B)). The microchannel height ($h_m$) is much larger than the particle diameter ($d_p$), so the continuous inflow of particles leads to the growth of a primary crystal layer followed by the formation of additional crystal layers underneath the primary layer (see Fig. 1(C)).

**Crystalline order:** To probe the arrangements of elements and defects in the crystalline lattice in a spatiotemporal manner, we employ bond-orientation angles (BOA). BOA can be determined for the individual particles by considering their adjoining neighbours using a Delaunay Mesh. For instance, the BOA for an element $i$ to its $n$ nearest neighbours, $\theta_i$ can be written as:

$$\theta_i = \sum_{k=1}^{n} \theta_{i,k},$$

where $\theta_{i,k}$ is the angle subtended from the line joining the centroids of the $i^{th}$ and the $k^{th}$ elements to a fixed/reference axis.

For an intuitive representation, the BOA can be delineated as a Voronoi plot (as in Fig. 2) with the Voronoi cells coloured according to the bond orientational angles. Different structures and orders within the lattice can then become apparent.

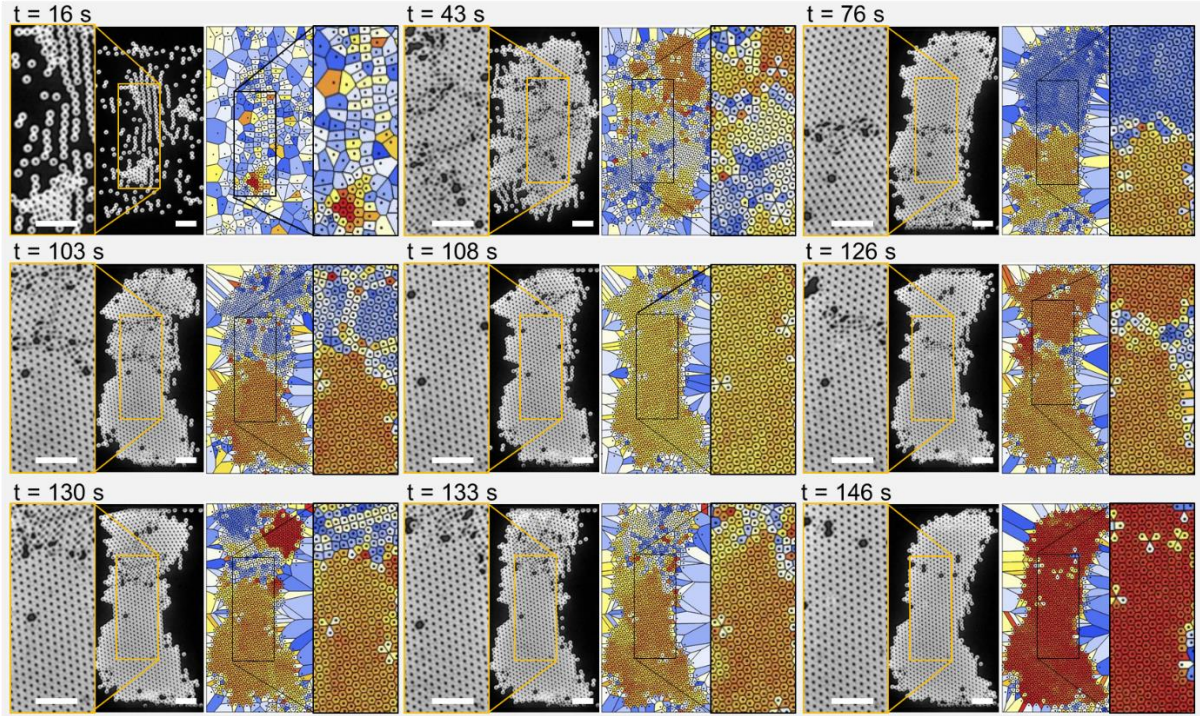

**Figure 2.** Experimental images (left panels) and Voronoi plots (right panels; the colors indicate the bond orientation angles; color homogeneity arises only for a single-crystal structure) are shown for a sequence of time steps with respect to the device actuation time. Initially, the self-assembly of the microparticles leads to the formation of small crystalline structures separated by cracks. These crystalline structures then join together as the cracks are removed, giving rise to a single-crystalline structure with a uniform crystal orientation. The sample flow rate was 250 μL/h. The device actuation frequency and input voltage were 72 MHz and 5.1 $V_{pp}$ respectively. The scale bars are 50 μm in length.

## Results and discussion

**Single-layer crystalline structure:** Figure 2 shows experimental images (left panels) and Voronoi plots (right panels) of a single-layer colloidal crystal (see Movie S1 in the SI). The particle sample solution was pumped through the microchannel at a flow rate of 250 μL/h while the IDT was actuated at a frequency of 72 MHz with an input voltage of 5.1 $V_{pp}$. The microparticles have a diameter of 10 μm and self-assemble mainly due to the influence of the TSAW-based direct ARF. However, note that after 16 s of device actuation, the particles started to concentrate in lines roughly parallel to the IDT because standing-wave-based pressure nodes have formed due to SAW reflection within the microchannel.[20,37] These particle lines join together to form small particle clusters that grow into larger grains as more particles flow into the particle trapping zone (at 43 s). The disordered colloidal particles (at 16 s) become relatively ordered crystalline structures (at 43 s) consisting of several segmented regions with various crystalline orientations and grain boundaries. The particle flow was stopped at this stage and a buffer solution was introduced with a similar flow rate through the second inlet into the microchannel, which ensured that the balance between the Stokes drag force and ARF was maintained so that the re-orientation and growth of the crystals into a larger single crystal could be studied. At 76 s, most of the cracks between the small crystal structures have been repaired: there are only two larger crystals present with a slightly different orientation and a dividing line in between them. The crack at the center (at 76 s) then moves upward giving rise to additional smaller cracks in the upper segment of the crystal as the crystalline structures reorient themselves (at 103 s). The self-repair of the cracks leads to a predominantly single-crystal structure with a few defects (particles missing from the crystal lattice), as confirmed by the uniform color of the Voronoi plots, which indicates the presence of a single crystal orientation (at 108 s). A new crack is then initiated just above the centerline of the microchannel, which again divides the main crystal into two parts (at 126 s). The self-repair of the crystal is initiated as the crystal structures re-orient themselves (at 130 s) and a crack moves outward from the larger crystal structure at the bottom (at 133 s). A single unified crystal with few defects and no cracks is again evident at 146 s. For additional results for the formation of single-layer crystals at different surfactant concentrations, see Fig. S1 and Movie S2 in the SI.

**Multiple-layer crystalline structure:** Figure 3 shows the self-assembly of microparticles to form multiple-layer colloidal crystals. A single-layer colloidal structure was readily formed within 1.2 min of the actuation of the TSAW-based device with an input voltage of 5.1 $V_{pp}$ resulting in a signal with a frequency of 72 MHz; the sample flow rate was 250 μL/h (see Fig. 3(A) and Movie S3 in the SI). The single-layer crystal evident in the inset (at 1.2 min) has a predominantly uniform orientation with only a few random defects (holes). The further inflow of particles leads to the formation of a secondary layer in the colloidal structure (depicted as a brighter region), and the defects present in the primary layer (at 1.2 min) are then filled by the additional particles (at 1.7 min). The inset shows the contrast between the single-layer and bi-layer colloidal structures. At 2.2 min, the secondary layer has expanded; the dark and bright shades correspond to multiple layers oriented differently with respect to the primary layer (see the inset). At 2.7 min, the crystal has grown to 450 μm in length; the extra bright regions close to the microchannel walls (see the inset) indicate the presence of multiple layers of colloidal particles that obscure the crystalline structure. Figure 3(B) shows contrasting experimental images for multiple-layer crystal growth for which the background noise has not been removed. For a sample flow rate of 500 μL/h, a predominantly bi-layer crystal has formed at 4 min, and the inset shows the contrast between the single-layer (bright) and bi-layer (dark) regions. As the crystal structure expands with the addition of new particles, the Stokes drag force acting on the crystal increases as well, so the input voltage was increased from 8.6 $V_{pp}$ to 10.4 $V_{pp}$ to impart an equivalent ARF and thus to maintain the position of the crystal (at 5 min). The input voltage was further increased to 12.2 $V_{pp}$ and 13.8 $V_{pp}$ at 7 min and 8 min respectively, as the crystal structure reached a length of 750 μm. For additional results with other surfactant concentrations, see Fig. S2 in the SI.

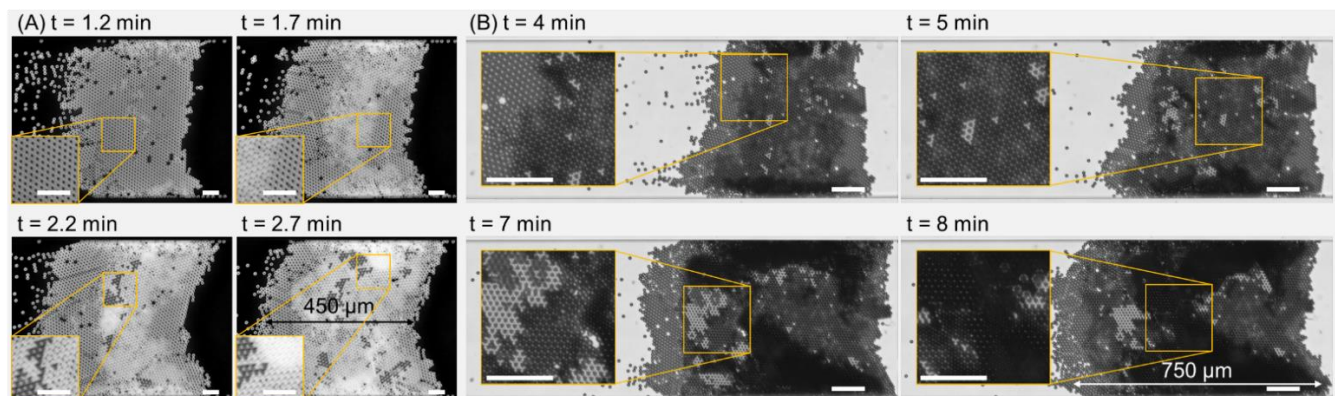

**Figure 3.** Experimental images, with background noise removed in (A) and without background noise removal in (B), of the self-assembly of multiple-layer crystalline structures. (A) The single-layer crystal at 1.2 min has become a multiple-layer crystal at 2.7 min. The sample flow rate was 250 μL/h and the input voltage was 5.1 $V_{pp}$ for a signal with a frequency of 72 MHz. Scale bars are 50 μm in length. (B) The predominantly bi-layer structure at 4 min has grown into a structure with more than two layers at 8 min, resulting in an overall crystal length of 750 μm. The sample flow rate was 500 μL/h, and the input voltage was gradually increased from 8.6 $V_{pp}$ to 13.8 $V_{pp}$. Scale bars are 100 μm in length.

**Input voltage and flow rate effects:** Figure 4 shows the effects of varying the input voltage and the buffer flow rate on the equilibrium position and shape of the crystal structure. Microparticles were pumped through the microchannel at 250 μL/h and a TSAW-based ARF with an input voltage of 4.2 $V_{pp}$ was applied to induce their self-assembly into a single-layer colloidal crystal (see Fig. 4(A)). The particle flow was replaced by a buffer flow (250 μL/h), and the input voltage was gradually increased up to 13.0 $V_{pp}$. At 6.5 and 8.6 $V_{pp}$, the crystalline structure has moved leftward while maintaining a rough rectangular shape. However, at 10.9 $V_{pp}$, the shape of the crystalline structure has become convex-concave, and this has become further exaggerated at 13.0 $V_{pp}$. For a constant flow rate, the Poiseuille flow profile remained constant; however, the acoustic forces increase with the voltage, resulting in a convex-concave shape at higher voltages for which the acoustic forces at the center of the microchannel dominate the Stokes drag force on the particles due to the Poiseuille flow.[16] The ARF profile is expected to have a uniform distribution close to the IDT across the microchannel width. However, it is also expected to have a lower amplitude farther away from the IDT near the microchannel walls because of the diffraction of the waves within the microchannel and their attenuation away from the IDT.[16] At 13.0 $V_{pp}$, the buffer flow rate was increased to 500 μL/h, which pushes the crystalline structure rightward (see Fig. 4(B)). When the flow rate was further increased to 1000 μL/h, the curvature of the convex-concave shape decreased as it regained a rectangular shape at 1500 μL/h. A concave-convex shape also arises at 2000 μL/h, and is particularly prominent at 2500 μL/h. At high flow rates, the flow velocity imparts a higher drag force on the particles at the center of microchannel and thus the crystal structure adopts a concave-convex shape. The effects of varying the input voltage and the buffer flow rate on the distance between the crystalline structure and the IDT are plotted in Figs. 4(C) and (D), respectively. Additional results confirming the above behavior are presented in Figs. S3-S5 (see the SI).

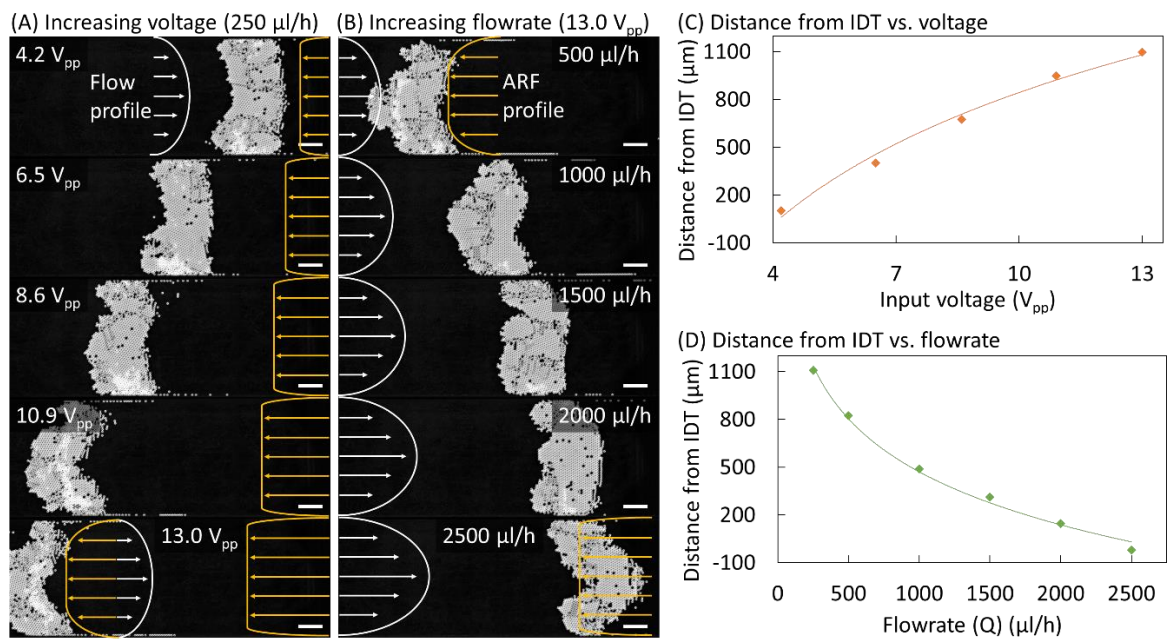

**Figure 4.** (A) For a constant sample flow rate of 250 μL/h the input voltage was increased from 4.2 $V_{pp}$ to 13.0 $V_{pp}$ to produce a signal with a frequency of 72 MHz. The crystal structure moves away from the IDT with increases in the input voltage. (B) The sample flow rate was increased from 500 μL/h to 2500 μL/h while the input voltage was kept constant at 13.0 $V_{pp}$. The crystal structure moves to the right as the sample flow rate increases. Scale bars are 100 μm in length. The distance of the crystal structure from the IDT was plotted against the variable input voltage (C) and the flow rate (D).

## Conclusions

We have demonstrated that an external acoustic force field can be used to trap continuously flowing microparticles so that they self-assemble into a crystalline structure: the acoustic radiation force due to the travelling surface acoustic waves is balanced by the Stokes drag force on the particles. The application of a continuous buffer solution flow, which balances the ARF, induces rearrangements of the particles that repair any cracks, dislocations, or defects in a single-layer crystal lattice. The defects (or particle holes) are filled by the additional incoming particles that stack beneath the primary crystal layer to form a multi-dimensional crystal lattice. The crystalline structure re-orients itself under the influence of the external force field. However, the movement of the crystal becomes restricted when the crystal size fills the gap between the microchannel walls, which locks the crystal motion. The propagation of TSAWs through the substrate and the coupling and diffraction of the acoustic waves within the microchannel lead to the evolution of the ARF profile away from the IDT. The balance between the ARF and the drag force induced by the Poiseuille flow usually results in a rectangular crystalline structure (see Fig. S5 in the SI); however, at higher input voltages and lower flow rates, a convex-concave structure results, and vice versa. The TSAW-based self-assembly of the microparticles leads to the formation of larger crystalline structures than are possible with a standing-acoustic-waves-based platform because acoustic pressure nodes have a low saturation limit.

## Acknowledgments

This work was supported by the National Research Foundation of Korea (NRF) (Grant No. 2018001483), the KUSTAR-KAIST Institute, and the Korea Polar Research Institute (KOPRI).

**Author contributions:** G. D. conceived the research. H. J. S. supervised the research. G. D., H. A., J. P. and M. A. designed and performed the experiments. G. D. A. H. and J. P. analyzed the results. All authors have contributed in preparing the manuscript.

**Corresponding Author:** Email: hjsung@kaist.ac.kr

# Supporting information

**Figure S1.** Self-assembly of microparticles suspended in a solution with 5% (A) and 10% (B) surfactant. The sample flowrate is 250 μl/h. The device actuation frequency and input voltage are 72 MHz and 5.1 $V_{pp}$, respectively. The scale bars are 50 μm in length.

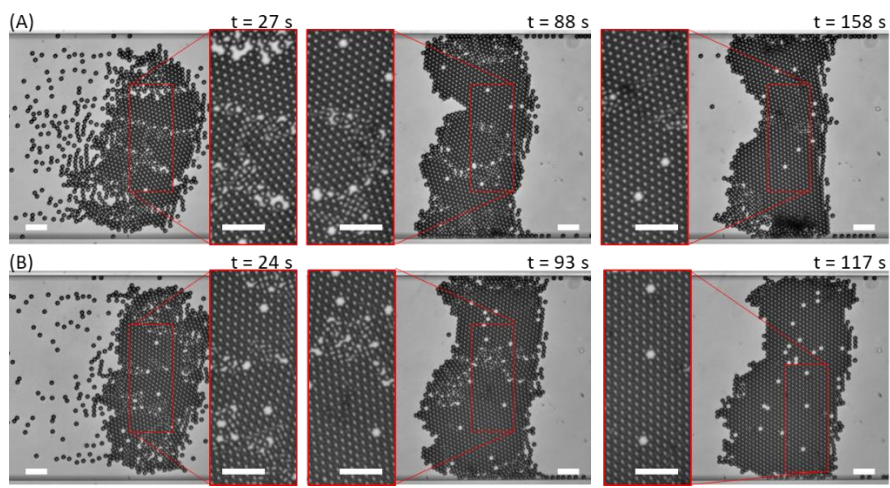

**Figure S2.** Experimental images for self-assembly of microparticles with a 5% surfactant into multiple layer crystalline structure. The sample flowrate is 250 μl/h. The device actuation frequency is 72 MHz, and the input voltage is increased from 6.9 $V_{pp}$ to 8.6 $V_{pp}$. Scale bars are 100 μm in length.

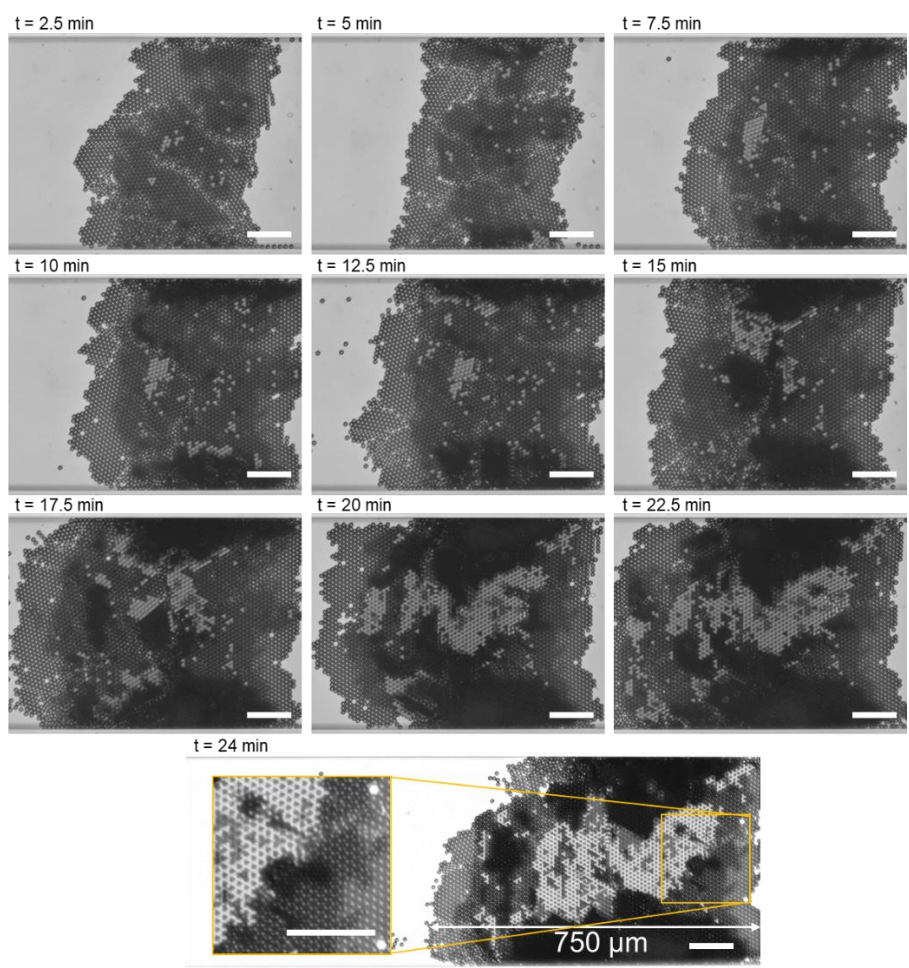

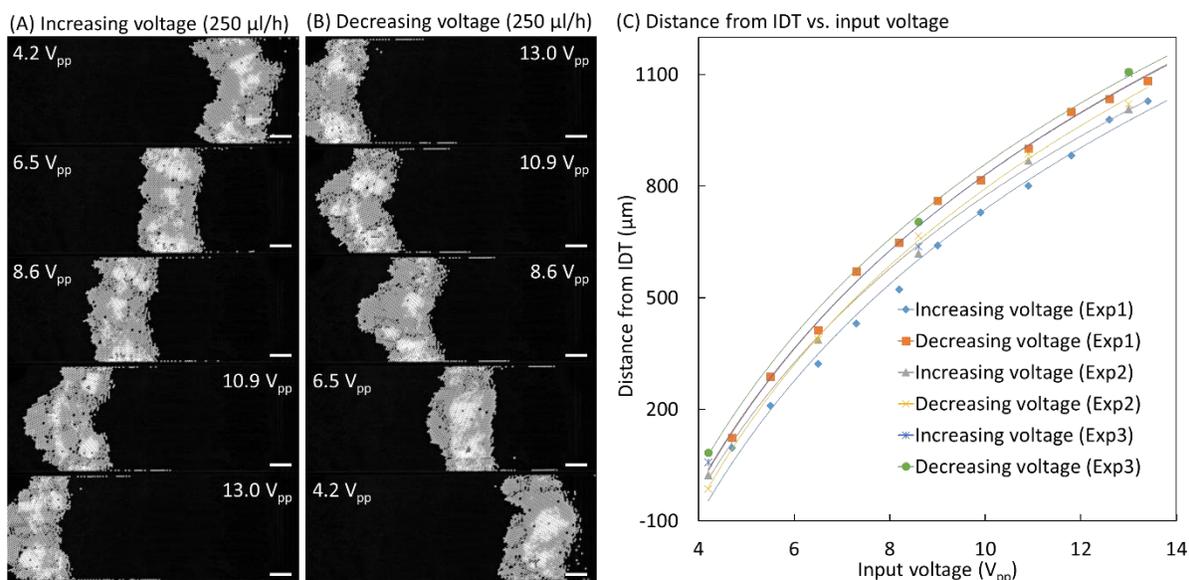

**Figure S3.** For a constant sample flowrate of 250 μl/h, the input voltage is increased (A) from 4.2 $V_{pp}$ to 13.0 $V_{pp}$ and then decreased (B) to 4.2 $V_{pp}$ for 72 MHz frequency signal as the colloidal structure moved leftward from the IDT and then back. Scale bars are 100 μm in length. (C) The colloidal structure distance from the IDT is plotted against variable input voltage for three different experiments.

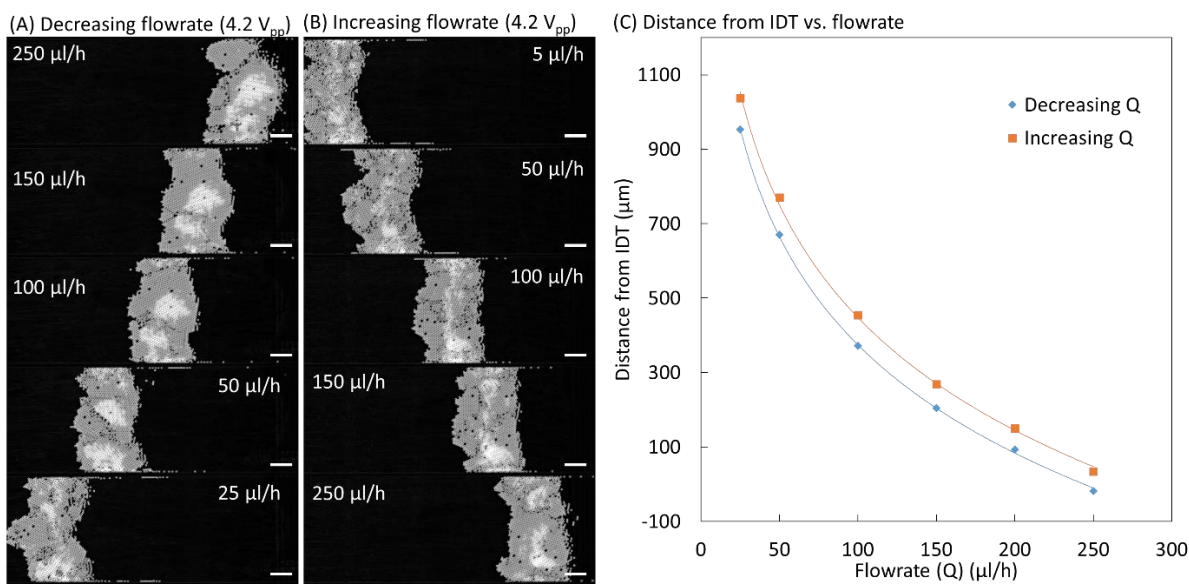

**Figure S4.** For a constant input voltage of 4.2 $V_{pp}$, the sample flowrate is decreased from 250 μl/h to 25 μl/h (A) and then increased to 250 μl/h (B) as the colloidal structure moved leftward from the IDT and then back. Scale bars are 100 μm in length. (C) The colloidal structure distance from the IDT is plotted against flowrate.

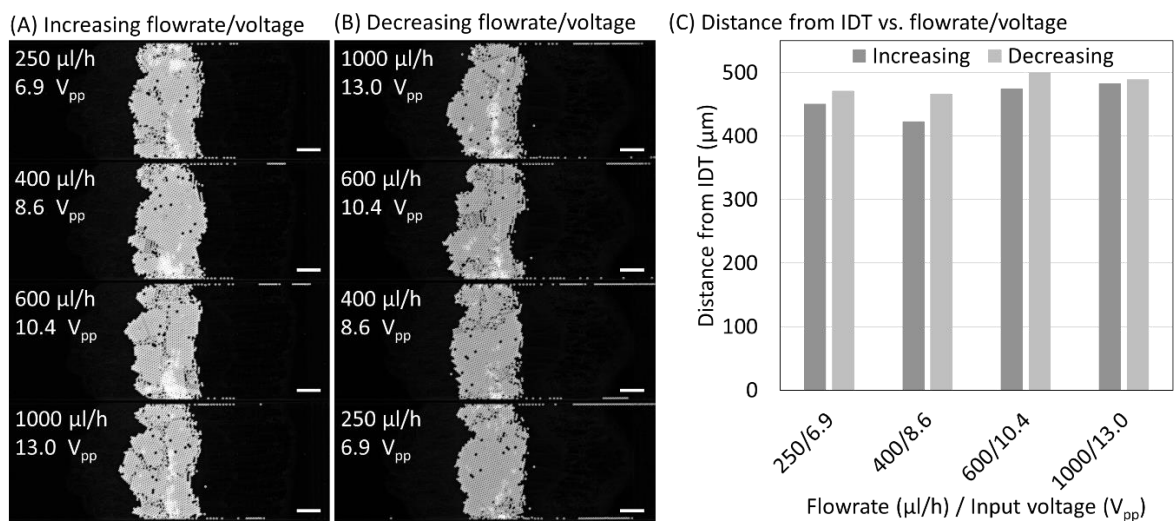

**Figure S5.** For an approximate consistent position of the colloidal structure inside the microchannel, a combination of input voltage and flowrate is respectively increased (A) and decreased (B) as the colloidal structure remained unmoved. Scale bars are 100 μm in length. (C) The colloidal structure distance from the IDT is plotted against flowrate and input voltage combinations.

## Movies captions
Movie 1: Single layer crystal formation. (see also Fig. 2)
Movie 2: Single layer crystal formation without the
background removed. The surfactant volume ratio is 10%.
(see also Fig. S1)
Movie 3: Multiple layer crystal formation. (see also Fig. 3)